\documentclass[aps,prl,twocolumn,superscriptaddress,showpacs]{revtex4-1}
\usepackage{amssymb,amsmath}
\usepackage{graphicx}
\usepackage[sort&compress]{natbib}
\usepackage{color}
\usepackage{multirow}
\usepackage{sidecap}
\usepackage{graphicx}
\usepackage{dcolumn}
\usepackage{bm}
\begin{document}
\title{Strong competition between orbital-ordering and itinerancy in a frustrated spinel vanadate}
\author{J. Ma}
\affiliation{Quantum Condensed Matter Division, Oak Ridge National Laboratory, Oak Ridge, Tennessee 37831, USA}
\author{J. H. Lee}
\affiliation{Materials Science and Technology Division, Oak Ridge National Laboratory, Oak Ridge, Tennessee, 37831, USA}
\author{S. E. Hahn}
\affiliation{Quantum Condensed Matter Division, Oak Ridge National Laboratory, Oak Ridge, Tennessee 37831, USA}
\author{Tao Hong}
\affiliation{Quantum Condensed Matter Division, Oak Ridge National Laboratory, Oak Ridge, Tennessee 37831, USA}
\author{H. B. Cao}
\affiliation{Quantum Condensed Matter Division, Oak Ridge National Laboratory, Oak Ridge, Tennessee 37831, USA}
\author{A. A. Aczel}
\affiliation{Quantum Condensed Matter Division, Oak Ridge National Laboratory, Oak Ridge, Tennessee 37831, USA}
\author{Z. L. Dun}
\affiliation{Department of Physics and Astronomy, University of Tennessee, Knoxville, Tennessee 37996, USA}
\author{M. B. Stone}
\affiliation{Quantum Condensed Matter Division, Oak Ridge National Laboratory, Oak Ridge, Tennessee 37831, USA}
\author{W. Tian}
\affiliation{Quantum Condensed Matter Division, Oak Ridge National Laboratory, Oak Ridge, Tennessee 37831, USA}
\author{Y. Qiu}
\affiliation{NIST Center for Neutron Research, Gaithersburg, Maryland 20899-6102, USA}
\affiliation{Department of Materials Science and Engineering, University of Maryland, College Park, Maryland 20742, USA}
\author{J. R. D. Copley}
\affiliation{NIST Center for Neutron Research, Gaithersburg, Maryland 20899-6102, USA}
\author{H. D. Zhou}
\affiliation{Department of Physics and Astronomy, University of Tennessee, Knoxville, Tennessee 37996, USA}
\author{R. S. Fishman}
\altaffiliation{author to whom correspondences should be addressed: fishmanrs@ornl.gov and matsudam@ornl.gov}
\affiliation{Materials Science and Technology Division, Oak Ridge National Laboratory, Oak Ridge, Tennessee, 37831, USA}
\author{M. Matsuda}
\affiliation{Quantum Condensed Matter Division, Oak Ridge National Laboratory, Oak Ridge, Tennessee 37831, USA}
\begin{abstract}

The crossover from localized- to itinerant-electron regimes in the geometrically-frustrated spinel system Mn$_{1-x}$Co$_x$V$_2$O$_4$ 
is explored by neutron-scattering measurements, first-principles caclulations, and spin models. 
At low Co doping, the orbital ordering (OO) of the localized V$^{3+}$ spins suppresses magnetic frustration 
by triggering a tetragonal distortion. 
At high Co doping levels, however, electronic itinerancy melts the OO and lessens the structural and magnetic anisotropies, thus increases the amount of geometric frustration for the V-site pyrochlore lattice.
Contrary to the predicted paramagentism induced by chemical pressure, 
the measured noncollinear spin states in the Co-rich region of the phase diagram provide a unique platform 
where localized spins and electronic itinerancy compete in a geometrically-frustrated spinel.

\end{abstract}

\pacs{61.05.fm, 75.10.Jm, 75.25.Dk, 75.30.Et}

\maketitle

The competition between localized and itinerant behavior triggers 
many intriguing phenomena such as metal-insulator transitions \cite{Imada}, colossal magnetoresistance \cite{CMR}, 
and superconductivity in heavy fermion \cite{Stewart} and Fe-based materials \cite{Kamihara}. 
Likewise, the transformation from itinerant to localized spins in geometrically-frustrated systems can create exotic phases 
by modifying the relationship between the spin, orbital, and lattice \cite{frust} degrees of freedom. 
Although the competing effects of localized and itinerant behavior on magnetic frustration 
have been rather extensively investigated on triangular and pyrochlore lattices, 
they have rarely been explored for the frustrated spinel $AB$$_2$O$_4$. 

Underlying the rich phase diagrams of spinels are the tuneable magnetic interactions between the $A$ and $B$ sites 
and the geometric frustration experienced by the $B$ sites on a pyrochlore lattice.
Spinel vanadates exhibit additional intriguing characteristics due to  
the orbital ordering (OO) \cite{Kismarahardja1, Nishiguchi, Lee, Wheeler, Garlea, Katsufuji, MacDougall1} of the 
partially-filled (3$d^2$) $B$-sites.
Because itinerancy interferes with OO \cite{Canosa,Kismarahardja1, Nishiguchi, Lee, Wheeler},
it enriches the complex interplay between the magnetic, orbital, and lattice degrees of freedom. 
But this complexity also makes it difficult to provide a detailed microscopic understanding 
of this system \cite{Tsunetsugu, Matteo, Sarkar,Chern,Whun, Katsufuji}. Through substitution of the A-site, we systematically study the competition between OO and itinerancy in the spinel vanadates. 
In particular, the modified spin, orbital, and lattice couplings in Mn$_{1-x}$Co$_x$V$_2$O$_4$ are used 
to reveal the competing effects of OO and itinerancy on the coupled magnetic and structural phase transitions \cite{Canosa, Kismarahardja, Kiswandhi}.

Elastic and inelastic neutron-scattering (INS) measurements are combined with first-principles calculations and spin models
to study single-crystals of Mn$_{1-x}$Co$_x$V$_2$O$_4$.  
Because Co$^{2+}$ is the smallest 2+ magnetic cation that can be introduced on the $A$ site, it
exerts strong chemical pressure. 
Previously, this chemical pressure was expected to induce itinerancy and consequent paramagnetism \cite{Canosa}. 
However, we find that the itinerancy driven by Co doping unexpectedly enhances both the para-to-collinear 
and collinear-to-noncollinear (CL to NC) transition temperatures, $T_{\rm CL}$ and $T_{\rm NC}$.
While the magnetic ground state of the V spins remains a two-in/two-out state 
throughout the entire doping range, that state has different origins in the itinerant (Co-rich) 
and localized (Co-poor) regimes.

At low Co doping, OO supresses frustration by triggering cubic-to-tetragonal structural 
and CL to NC (two-in/two-out) magnetic transitions at $T_S = T_{NC}$.
However, at high Co doping, itinerancy weakens the structural transition ($ T_S < T_{NC}$) 
by melting the OO and revives the suppressed frustration. 
Despite the complete disappearance of OO and of the structural transition, 
the novel two-in/two-out is stabilized 
by the revivived frustration and the enhanced Co-V exchange at an even higher temperature than at low doping levels. 

\begin{figure}
 \centering
  \includegraphics[width=0.465 \textwidth]{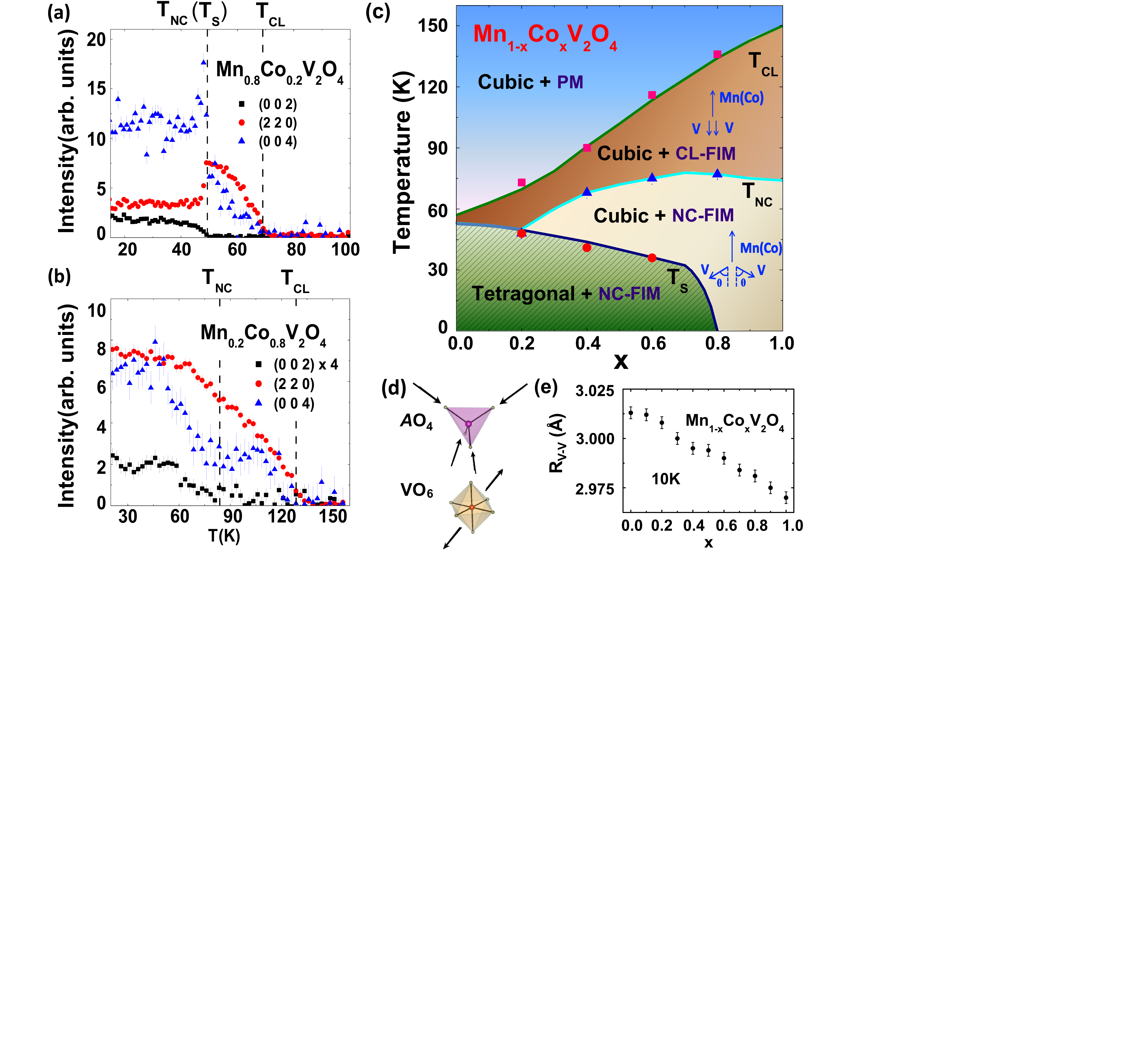}\\
\caption{ (color online) Temperature dependence of the Bragg peaks, (002) (squares), (220) (circles), and (004) (triangles)
in (a) Mn$_{0.8}$Co$_{0.2}$V$_2$O$_4$ and (b) Mn$_{0.2}$Co$_{0.8}$V$_2$O$_4$. 
  The background has been subtracted. (c)  The temperature versus Co-doping content ($x$) phase diagram. 
  The para-to-CL magnetic transition temperature $T_{CL}$, the CL-NC ferrimagnetic phase transition temperature $T_{NC}$, 
  and the cubic-to-tetragonal lattice transition temperature $T_S$ are determined from the magnetic susceptibility/heat capacity (solid lines) 
  and neutron-scattering experiments (square, triangle, and circle for the transitions). (d) $A$O$_4$ ($A$=Mn$^{2+}$/Co$^{2+}$) tetrahedron and VO$_6$ octahedron. 
  (e) The $x$ dependence of the V$^{3+}$-V$^{3+}$ distance, $R_{V-V}$, at 10 K (black dots). }
  \label{diff_gap}
\end{figure}

Neutron-diffraction experiments were performed at the four-circle diffractometer (HB-3A) and the triple-axis spectrometer (HB-1A) at the High Flux 
Isotope Reactor (HFIR) of the Oak Ridge National Laboratory (ORNL). The data were refined by the Rietveld method using FULLPROF \cite{Juan}.  
INS data were collected utilizing the thermal (HB-1) and cold (CG-4C) triple-axis spectrometers at HFIR, ORNL, with fixed final energies of 
14.7 meV at HB-1 and 5 or 3.5 meV at CG-4C, respectively; and the time-of-flight (tof) spectrometers, DCS, at National Institute of Standards and Technology 
(NIST) \cite{Copley} and SEQUOIA, at the Spallation Neutron Source (SNS) with fixed incident energies of 25.25 and 30 meV, respectively. 
The tof data were analyzed with DAVE \cite{Azuah}. Error bars in the figures represent one standard deviation.

The Co-doping dependence of the crystal and magnetic structures and OO was determined by single-crystal neutron-diffraction measurements. 
Figures~\ref{diff_gap}(a) and (b) show the temperature dependence of the (002), (220), and (004) Bragg peaks 
for Mn$_{1-x}$Co$_x$V$_2$O$_4$ ($x$=0.2 and 0.8).
A ferrimagnetic (FIM) signal develops below $T_{CL}$ at the symmetry-allowed Bragg positions (220) and (004). 
While the (002) peak is forbidden by symmetry, the observed scattering intensity below $T_{NC}$ signals
the formation of an antiferromagnetic (AFM) spin structure in the $ab$-plane. The onset of the (002) 
magnetic reflection marks the CL-NC magnetic transition at $T_{NC}$.
The (004) reflection, which also increases in intensity below $T_{NC}$, provides a measure of both magnetic transitions.
For $x$=0.2, the intensities of the (220) and (004) Bragg peaks rise at $\sim$70 K ($T_{CL}$)
due to the para-to-CL magnetic transition. The (220) peak drops sharply at $\sim$50 K ($T_S$=$T_{NC}$) due to
the cubic-to-tetragonal structural transition. Hence, $T_{NC}$ coincides with $T_S$ through $x$$\approx$0.2 \cite{Garlea}. 
For $x>0.2$, the two transitions separate with $T_S < T_{NC}$.
At $x$=0.8, the structural transition disappears, while two magnetic transitions are observed with $T_{CL}$$\sim$150 K and $T_{NC}$$\sim$80 K. In high Co-doped compounds, x-ray diffraction and heat capacity measurements also suggest the absence of a structural transition \cite{Kiswandhi}.

To summarize the results as shown in the phase diagram in Fig.~\ref{diff_gap}(c):
we find for i) $x$$\le$0.2: a para-to-CL magnetic transition at $T_{CL}$ and
a cubic-to-tetragonal structural transition that coincides with the CL-NC transition at $T_S=T_{NC}<T_{CL}$;
ii) 0.2$<$$x$$<$0.8: the CL-NC and cubic-to-tetragonal transitions are decoupled with
$T_S<T_{NC}<T_{CL}$;  iii) $x$$\ge$0.8:  no structural transition is observed down to 5 K but
two magnetic phase transitions appear with $T_{NC}<T_{CL}$. Both $T_{CL}$ and $T_{NC}$ increase while $T_S$ gradually 
decreases with Co$^{2+}$ doping. The detailed measurements used to map the phase diagram are provided in Supplementary Information(SI) A and B \cite{SI}. 

The microscopic effect of Co doping can be understood by considering the structures of the $A$O$_4$ tetrahedra and VO$_6$ octahedra.
In the high-temperature cubic phase ($Fd\bar{3}m$), the interior angles of the $A$O$_4$ tetrahedra are $\angle$O-$A$-O=109.7$^\circ$.
With Co doping, the $A$-O bond length decreases from 2.041 \AA\ ($x$=0) to 1.984 \AA\ ($x$=0.8), thereby applying
chemical pressure along the $A$-O direction,  Fig.~\ref{diff_gap}(d).  Each VO$_6$ octahedra stretches along the 
$<$111$>$ direction, producing the local trigonal distortion shown in Fig.~\ref{diff_gap}(d).  Due to this distortion, 
the twelve O-V-O interior angles in the VO$_6$ octahedra split away from 90$^\circ$ into two different angles. 
With Co$^{2+}$ doping, the difference between the two O-V-O angles decreases from 12.3(2)$^\circ$ ($x$=0) to 10.0(2)$^\circ$ ($x$=0.8). 
Simultaneously, the V-O bond length shrinks from 2.023(1) ($x$=0) to 2.012(1)\AA\ ($x$=0.8) 
and the V-V bond length ($R_{V-V}$) shirnks from 3.013 to 2.975 \AA, which increases both chemical 
pressure and structural isotropy. 

\begin{figure*}
 \centering
 \includegraphics[width=0.95 \textwidth]{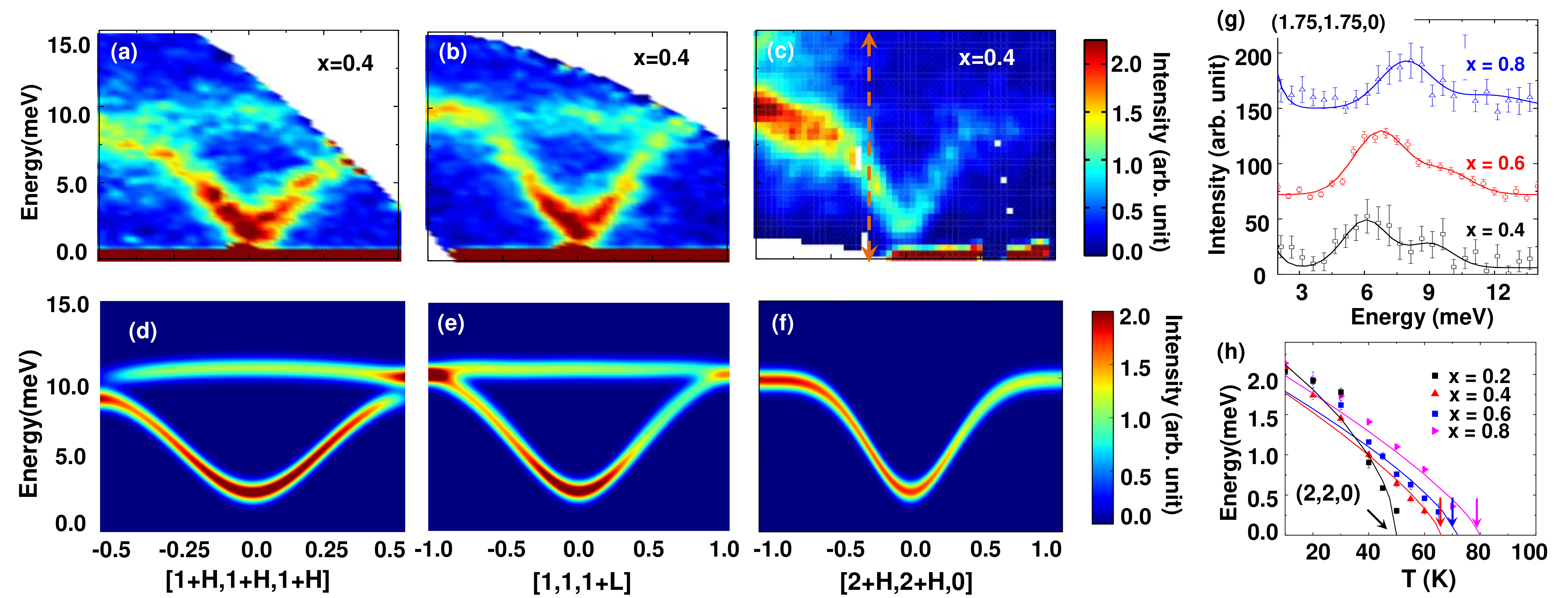}
\caption{ (color online) (a)-(c) INS results for the magnetic excitations of Mn$_{0.6}$Co$_{0.4}$V$_{2}$O$_{4}$ at 8 K. (d)-(f) The calculated excitation using the Hamiltonian, Eq. (1). 
The arrow line in (c) represents the position of the constant-$Q$ cut. (g) The constant-$Q$ cuts at (1.75 1.75 0) along [H H 0] direction in Mn$_{1-x}$Co$_{x}$V$_{2}$O$_{4}$ measured at 8K, $x$=0.4, 0.6 and 0.8.  The curves in (g) are Gaussian fits and guides to the eye.  Note that the low-energy spin-wave branch hardens with Co-doping. (h) The spin-wave energy gap at the magnetic zone center (2 2 0) in Mn$_{1-x}$Co$_{x}$V$_{2}$O$_{4}$. The curves in (h) are power-law fits.}
  \label{structure}
\end{figure*}

The structural phase diagram reflects the evolution of these bond length and angle parameters with Co doping. 
Although the crystal space group changes from $Fd\bar{3}m$ (cubic) to $I4_1/a$ (tetragonal) 
with decreasing temperature when $x$$<$0.8, it remains $Fd\bar{3}m$ (cubic) down to the lowest temperature studied when $x$$\ge$0.8. 
By contrast, most spinel vanadates exhibit structural transitions with decreasing 
temperature\cite{Kismarahardja1, Nishiguchi, Lee, Wheeler, Garlea, Katsufuji, MacDougall1}, so the behavior of the $x$$\ge$ 0.8 samples is anomalous.

On the other hand, magnetic structures is tightly coupled to the crystal structure and very sensitive to the Co doping level. 
Below $T_{CL}$, the Mn/Co moments are aligned parallel to the $c$ axis.
Above $T_{NC}$, the V$^{3+}$ moments point along the $c$ axis and antiparallel to the Mn/Co moments.
Below $T_{NC}$, the V$^{3+}$ moments form the two-in/two-out configuration observed previously in MnV$_2$O$_4$ \cite{Magee} and 
FeV$_2$O$_4$ \cite{MacDougall}. The canting of the V$^{3+}$ moments away from the $c$ axis starts below 70 K and reaches 
22.1(1.8)$^\circ$ at 10 K for $x$=0.8, smaller than 35.7(1.5)$^\circ$ 
and 36.2(1.5)$^\circ$ for $x=$0.0 and 0.2 at 10 K, respectively. 
Meanwhile, the V$^{3+}$ ordered moment initially increases from 0.95(4)$\mu_B$ ($x$=0.0) to 1.03(7)$\mu_B$ ($x$=0.2), then decreases to 
0.61(3)$\mu_B$ ($x$=0.8) at 10 K. The enhancement of the V$^{3+}$ moment from $x=$ 0.0 to 0.2 clearly reflects the 
reduced orbital moment associated with Co doping.  Contrary to the prediction 
that the small Co$^{2+}$ cation
triggers paramagnetism \cite{Canosa}, we find that the ordered V magnetic moment does not disappear for Co-rich ($x \ge 0.8$) compounds
and their ordering temperatures ($T_{CL}$, $T_{NC}$) even increase with doping, as shown in Fig.~\ref{diff_gap}(c).

\begin{figure*}
\centering
\includegraphics[width=0.95\textwidth]{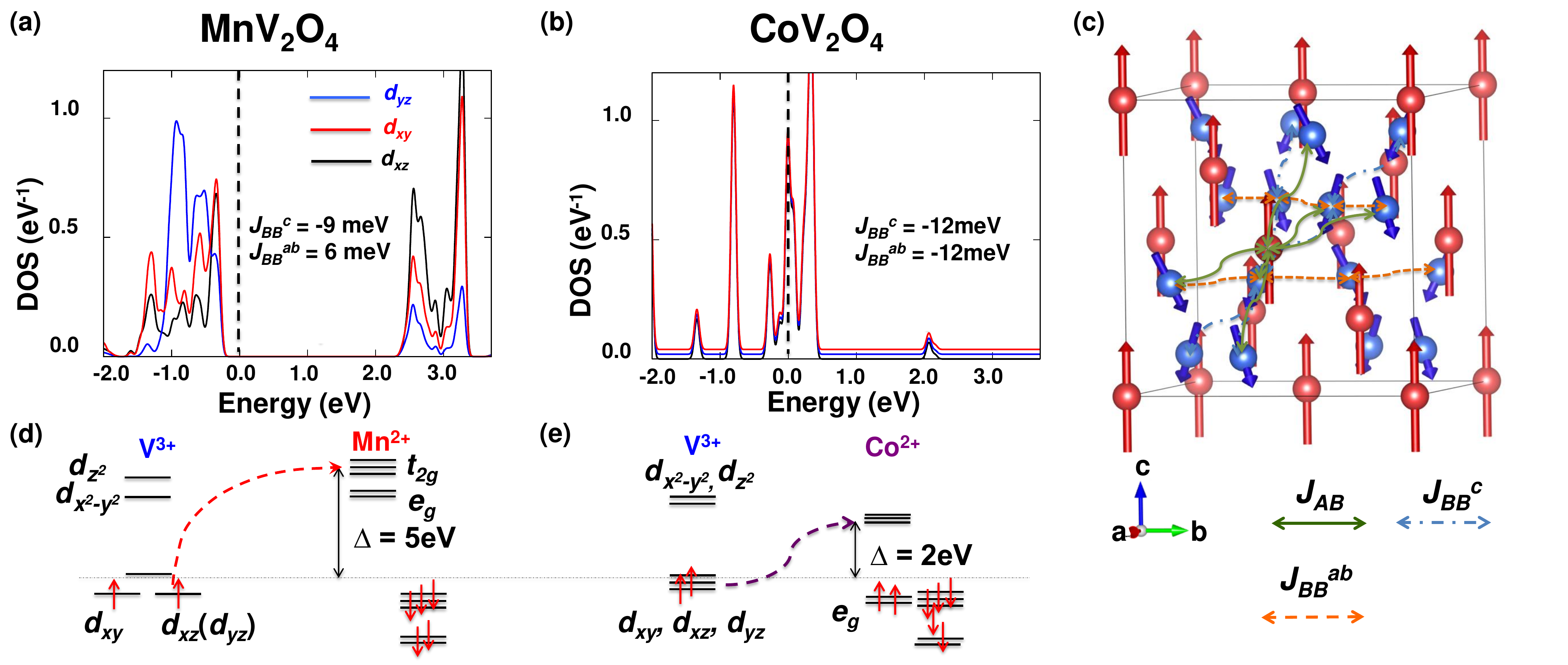}\\
\caption{(color online) (a) and (b) Density-of-states for AFM MnV$_2$O$_4$ and CoV$_2$O$_4$, respectively. 
(c) NC state of Mn$_{1-x}$Co$_x$V$_2$O$_4$. $J_{AB}$, $J^{ab}_{BB}$, and $J^{c}_{BB}$ 
are the exchange interactions between nearest-neighbor sites. (d) and (e) Orbital energies estimated from DFT calculations. 
  $|$$J_{A-{\rm V}}$$|$ is inversely proportional to the energy gap, $\Delta$.
 Only up-spin energy levels are shown for V$^{3+}$ for simplicity.}
  \label{phase}
\end{figure*}

To confirm the itinerancy-induced origin of the NC states, the exchange interactions and anisotropies were evaluated from the 
INS spectra of spin-wave excitations. Since the interaction between the $A^{2+}$ ions is known to be small \cite{Chung, Nanguneri,Suzuki}, 
the interactions between the 
$A^{2+}$ and V$^{3+}$ ions can be estimated from the dispersion of the low-energy acoustic mode.
For $x$=0.4, the measured dispersions along the $<$110$>$ direction are plotted in Fig.~\ref{structure}(a).
Notice that two magnetic modes have been observed, as shown in  Fig.~\ref{structure}(c), which is the same as in MnV$_2$O$_4$ \cite{Chung}, and the spin-wave velocities increase with Co doping. Measured at the (220) 
zone center, the spin-wave gap is plotted as a function of temperature and doping in Fig.~\ref{structure}(d). 
As in MnV$_2$O$_4$ \cite{Magee} and FeV$_2$O$_4$ \cite{MacDougall}, the spin-wave gap below $T_{NC}$ is 
produced by easy-axis anisotropy along the cubic diagonal of each V tetrahedron, which also cants the V$^{3+}$ moments 
away from the $c$ axis. Intriguingly, this spin-wave gap is almost independent of Co-doping, Fig.~\ref{structure}(d). 
Ignoring $J_{BB}^{ab}$ and $J_{BB}^c$, the spin-wave gap would be proportional to $\sqrt{J_{AB}\!\times\!D_{B}}$.  
We conclude that a roughly constant spin-wave gap of about 2 meV is maintained by the balance between the enhanced $J_{AB}$ and the 
suppressed $D_{B}$ associated with the itinerancy of the V$^{3+}$ ions. In particular, the suppressed $D_B$ promotes
frustration in the spinel structure \cite{Ising}.

Spin-wave theory (SWT) was used to understand the microscopic origin of the itinerancy-driven NC states in the absence of OO. These calculations were based on the Hamiltonian with six inequivalent sublattices,  
\begin{align}
\label{ModelHamiltonian}
H=    & - J_{AB} \sum \limits_{\left(p,q \right)\left(i,j,k,l\right)} \left(\boldsymbol{S}_p + \boldsymbol{S}_q \right) \cdot 
        (\boldsymbol{S}_i + \boldsymbol{S}_j + \boldsymbol{S}_k + \boldsymbol{S}_l) \nonumber \\
     & - J_{BB}^{ab} \left( \sum \limits_{i,j}  \boldsymbol{S}_i \cdot \boldsymbol{S}_j 
          + \sum \limits_{k,l} \boldsymbol{S}_k \cdot \boldsymbol{S}_l \right) \nonumber  \\
     & -J_{BB}^c \sum \limits_{(i,j)(k,l)} (\boldsymbol{S}_i +\boldsymbol{S}_j) \cdot \left( \boldsymbol{S}_k + \boldsymbol{S}_l \right) \nonumber \\
     & + D_A \sum \limits_{r=p,q} \left( \hat{z} \cdot \boldsymbol{S}_r \right)^2 
      + D_B \sum \limits_{s=i,j,k,l} \left( \hat{u_s} \cdot \boldsymbol{S}_s \right)^2 
\end{align}
The inequivalent $A$-sites are given by subscripts $p$ and $q$, and the inequivalent $B$-sites are given by subscripts $i$, $j$, $k$ and $l$. 
Isotropic exchange constants ($J_{AB}$, $J_{BB}^{ab}$, and $J_{BB}^c$) describe nearest-neighbor interactions, Fig.~\ref{phase}(c). 
For the $A$-site spins, the easy-axis anisotropy, $D_{A}$,  is along the $c$-axis while for the $B$-site spins, the easy-axis anisotropy, $D_{B}$, is along the local $<$111$>$ direction $\left( \hat{u_s} \right)$. A  range of values for $S_B$ and $J_{BB}^{ab}$ produces fits of similar quality, (SI C.2) \cite{SI}. Parameters best describing the experimental data for Mn$_{0.6}$Co$_{0.4}$V$_2$O$_4$ with $S_A$ = 4.2 $\mu_B$, $S_B$ = 1.4 $\mu_B$ and $J_{BB}^{ab}$ = $-$8.0 meV 
were $J_{AB}$ = $-$1.8 meV, $J_{BB}^c$=1.1 meV,
$D_B$= $-$9.1 meV and $D_A$ = 0.4 meV. 

The simulated dispersions of Mn$_{0.6}$Co$_{0.4}$V$_2$O$_4$ agree well with the measurements, Fig.~\ref{structure}.  With Co doping, we fix the $B$-site (V) moment while 
lowering the $A$-site (Mn/Co) moment. 
As a result, the exchange  $J_{AB}$ for Mn$_{0.6}$Co$_{0.4}$V$_2$O$_4$ 
is stronger than for MnV$_2$O$_4$ (SI C.2) \cite{SI}. 
By inducing electronic itinerancy, density-funtional theory (DFT) indicates that Co doping also strengthens 
both the structural ($c$ $\sim$ $a$) and magnetic ($J_{BB}^{ab}\sim J_{BB}^c$) isotropies, 
as shown in Figs.~\ref{phase}(a) and (b). 
If the $J_{AB}$ interactions were not enhanced by Co doping, the remanent magnetic 
anisotropies along the diagonals of the V tetrahedra would transform the V spin state into an all-in/all-out structure.  
Due to the enhanced $J_{AB}$, however, the ground state of the V spins 
remains the same isosymmetric two-in/two-out state found for small Co doping.

With the orbital energies of both the $A$ and $B$ ions estimated from DFT (SI. D) \cite{SI}, the origin of the enhanced $J_{AB}$ is 
explained in Fig. \ref{phase}(d) and (e).  The large energy difference ($\sim$ 5 eV) between the occupied V and Mn $d$ states 
weakens the exchange between Mn and V. By filling the $e_g$ level, Co doping significantly lowers the $t_{2g}$ level and enhances 
the exchange interaction between Co and V. DFT calculations reveal that the AFM $J_{AB}$ is significantly enhanced in 
CoV$_2$O$_4$ ($-$2.5 meV) compared to MnV$_2$O$_4$ ($-$1.2 meV). Although the V electrons are delocalized by Co doping, the enhanced 
$J_{AB}$ causes $T_{CL}$ to grow.  Further, the enhanced magnetic exchange isotropy ($J_{BB}^{ab}$ $\sim$ $J_{BB}^c$) driven by 
orbital quenching (Fig~\ref{phase}(b)) stabilizes the isosymmetric NC phase and raises $T_{NC}$. Therefore, the induced itinerancy 
strengthens both the CL and NC phases even without OO.

Induced itinerancy is closely related to $R_{V-V}$. At 10 K, Fig. \ref{diff_gap}(e) shows that $R_{V-V}$ 
remains almost constant up until $x=0.2$, then begins to decrease. Based on our DFT calculations, the 
shorter $R_{V-V}$ induces itinerant electronic behavior, as shown in Figs.~\ref{phase}(a) and (b), 
thereby suppressing OO. Due to the disappearance of OO by the itineracy, $T_S$ falls with Co doping.

\begin{figure}
 \centering
 \includegraphics[width=0.4 \textwidth]{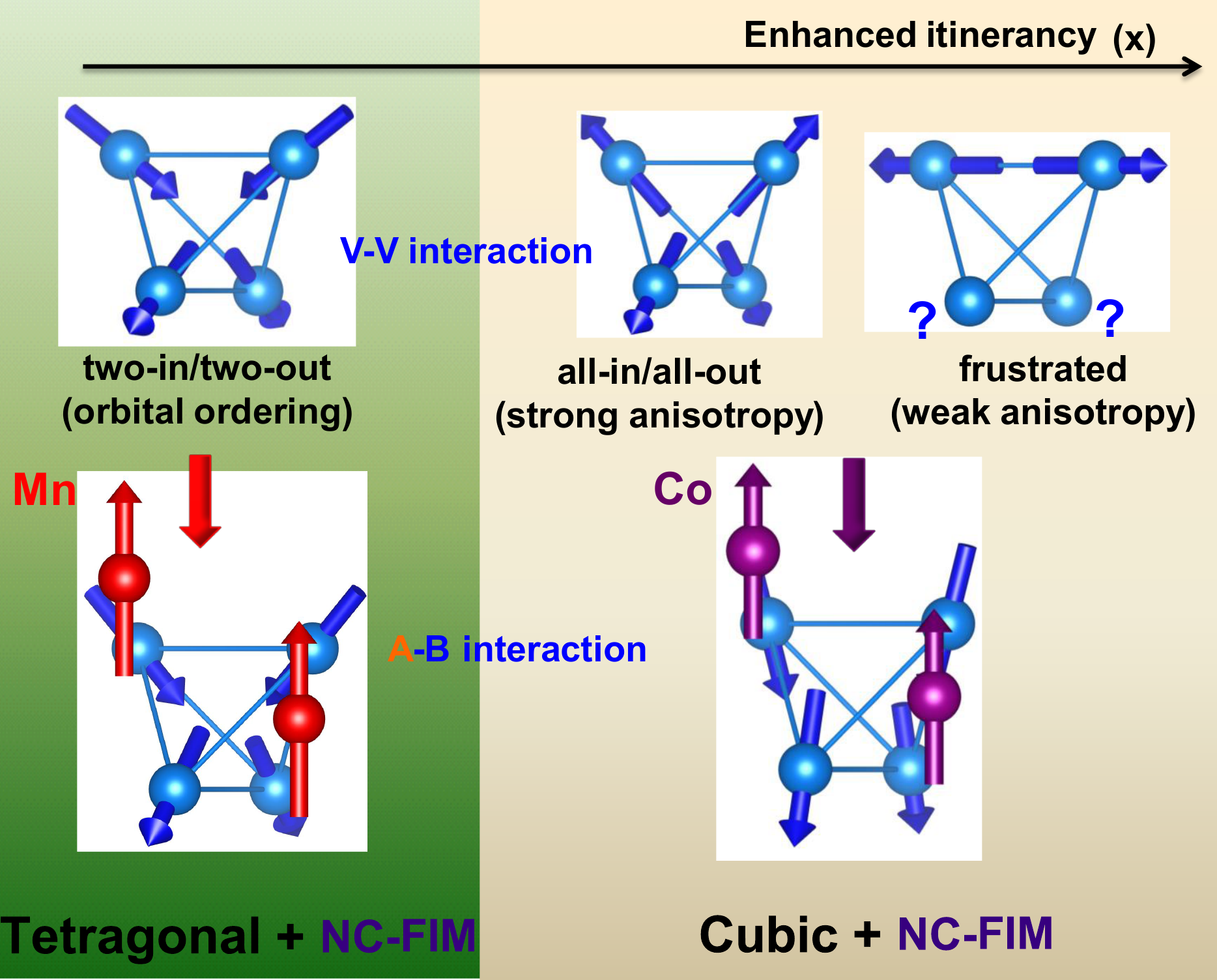}
\caption{ (color online) The hierarchical magnetic states with V-V and V-Mn/Co interactions and 
the distinct origins of isosymmetric phase transition with Co-doping ($x)$.} 
  \label{disp}
\end{figure}

As outlined in Fig.~\ref{disp}, the isosymmetric NC states have distinct origins for low and high $x$. 
For low Co-doping, the OO of the V ions relieves the magnetic frustration by triggering a tetragonal structure transition (c/a$<$1) 
and induces the two-in/two-out spin state.
The Mn-V interactions only increase the canting angle while maintaining the two-in/two-out. 
If only the isotropic V-V interactions and the remanent local V anisotropy were considered, Co doping would produce an 
all-in/all-out spin state.
However, the strong AFM $J_{AB}$ between the Co and V sites stabilizes the observed two-in/two-out state in the high Co-doping compounds.
Perturbations such as pressure may further strengthen the electronic itinerancy, weakening the remanent anisotropy and enhancing 
the magnetic frustration.

To summarize, Mn$_{1-x}$Co$_x$V$_2$O$_4$ exhibits a rich phase diagram due to the crossover from localized to itinerant electronic regimes.
The crystallographic and magnetic structures of compounds with low and high Co doping levels have different physical origins.
At low Co doping, OO triggers a cubic-to-tetragonal lattice distortion, accompanied by a CL-to-NC magnetic transition.
Co doping contracts $R_{V-V}$, enhances the electronic itineracy, and revives the magnetic 
frustration of the pryochlore lattice by weakening the magnetic and structural anisotropies.
With further Co doping, OO completely disappears and the magnetic ordering temperatures $T_C$ and $T_{NC}$ are driven higher 
by the enhanced exchange interaction $J_{AB}$.
Since CoV$_2$O$_4$ is located at the crossover between localized and itinerant electron behavior, 
external pressure may further strengthen itinerancy and magnetic isotropy,
enhance geometric frustration, and produce other exotic behavior. 
The present results provide a microscopic picture for the competition between OO and electronic itinerancy in this frustrated spinel series 
and suggest a new methodology for studying competing effects with multiple order parameters.

The research at HFIR and SNS, ORNL, were sponsored by the Scientific User Facilities Division (J.M., J.H.L, S.E.H., T.H., H.B.C., A.A.A., M.S., W.T., M.M.) 
and Materials Science and Engineering Division (J.H.L., R.F.), 
Office of Basic Energy Sciences, US Department of Energy. S.E.H. acknowledges support 
by the Laboratory's Director's fund, ORNL. Z.L.D and H.D.Z. thank the support 
from NSF-DMR through award DMR-1350002. 
Work at NIST is supported in part by the National Science Foundation under Agreement No. DMR-0944772. 
The authors acknowledge valuable discussions with S. Okamoto and G. MacDougall.

\bibliographystyle{apsrev}

\end{document}